# Noninvasive optical estimation of CSF thickness
# for brain-atrophy monitoring

Daniele Ancora[1,2]*, Lina Qiu[3], Giannis Zacharakis[1], Lorenzo Spinelli[4],
Alessandro Torricelli[3,4] and Antonio Pifferi[3,4]

[1]Institute of Electronic Structure and Laser, Foundation for Research and Technology - Hellas, Heraklion, Greece.
[2]Department of Materials Science and Technology, University of Crete, Heraklion, Greece.
[3]Dipartimento di Fisica, Politecnico di Milano, Milan, Italy
[4]Istituto di Fotonica e Nanotecnologie, Consiglio Nazionale delle Ricerche, Milan, Italy

*Corresponding author, email: daniele@iesl.forth.gr, address: N. Plastira 100, 70013 Heraklion (Greece)

**ABSTRACT (200 words)**

Dementia disorders are increasingly becoming sources of a broad range of problems, strongly interfering with normal daily tasks of a growing number of individuals. Such neurodegenerative diseases are often accompanied with progressive brain atrophy that, at late stages, leads to drastically reduced brain dimensions. At the moment, this structural involution can be followed with XCT or MRI measurements that share numerous disadvantages in terms of usability, invasiveness and costs. In this work, we aim to retrieve information concerning the brain atrophy stage and its evolution, proposing a novel approach based on non-invasive time-resolved Near Infra-Red (tr-NIR) measurements. For this purpose, we created a set of human-head atlases, in which we eroded the brain as it would happen in a clinical brain-atrophy progression. With these realistic meshes, we reproduced a longitudinal tr-NIR study exploiting a Monte-Carlo photon propagation algorithm to model the varying cerebral spinal fluid (CSF). The study of the time-resolved reflectance curve at late photon arrival times exhibited peculiar slope-changes upon CSF layer increase that were confirmed under several measurement conditions. The performance of the technique suggests good sensitivity to CSF variation, useful for a fast and non-invasive observation of the dementia progression.

# INTRODUCTION

In modern societies, demographic changes related to population ageing are already exhibiting dramatic impact from a personal, societal and economic point of view. Invalidating diseases slowly and inexorably push at the edge of the society the sufferers and their families, allowing little possibilities to maintain decent life expectations. Medical science and closely related research fields are currently pursuing a broad counter attack to such health issues, aiming at the full comprehension and cure of the sickness. A clear understanding of the disease generation and its evolution are the key point for reducing, alleviating or arresting the progression of the illness, thus helping the individual to return back to a normal-life condition.

In this scenario, generic Dementia symptoms –that progressively impair mental abilities– are one of the most difficult enemies to tackle. Accounting for 60-80% of the total cases diagnosed, *Alzheimer's disease* (AD) is the most common type of dementia, followed by *vascular dementia* (VD), *dementia with Lewy bodies* (DLB), *Parkinson's disease* (PD), *normal pressure hydrocefalus* (NPH), *frontotemporal dementia* (FTD) and other rarer conditions (source: www.alz.org [1] and [2]). At the moment, this class of diseases have no cure and neither a clear comprehension of the phenomena behind their evolution. Moreover, there are no validated ways to diagnose them –especially at early stages– and neither a way to stop or prevent the disease progression. Several ongoing researches have selected potential biomarkers which exhibit good ability to indicate early stage of the dementia disorders. Brain imaging [3] [4] [5], amyloid-PET imaging [6], presence of proteins in cerebral spinal fluid (CSF) [7] [8] [9] or in blood [10] and genetic risk profiling [11] are exceptionally good candidates, but at the moment they lack of general validation. Almost all of them rely on invasive procedures to evaluate the markers: Contrast agent injection [12], CSF sampling from the spinal cord [13] or blood sampling [5] [14] by using syringes are the typical diagnostic options. Not only they are invasive, but they also rely on XCT, PET, MRI measurements or post-sampling chemical analysis: in general these procedures are slow in measuring/interpreting the results and definitely very expensive for the healthcare system.

There is –without any doubt– a clinical need for a fast and easy way to detect, monitor and follow in time the disorders helping Dementia research to reach a wider number of patients. Apart from monitoring symptoms of memory loss and difficulties in performing normal every-day tasks, one of the less invasive way to follow the dementia progression is the MRI-estimation of the cerebral atrophy. In fact, brain volume reduction is associated with many kind of dementia and in particular with AD [3] [4], NPH and FTD [15]. In turns, the disease progression implies the increase of CSF volume circulating in specific regions of the braincase, to compensate the reduction of the brain. Such fluid plays the role of nutrient vector and mechanical cushion for the brain, having the potential to be one of the most reliable biomarker for the diagnosis of dementia disorders. In particular, the analysis of the CSF can report traces of beta-amyloid plaques which develop on the brain external cortex when the AD progresses, thus leading to a quite reliable tool for the diagnosis of this disease. Interestingly this fluid –constituted by 99% of water– results to be optically transparent in the visible and near-infrared range, having almost negligible absorption and scattering coefficients. In these terms, a generic dementia development could be seen as a volume-increase of non-scattering regions within the head that would strongly affects the photon diffusion within the tissues, as already proven by several works on the fields [16] [17] [9].

In our work, we examine the possibility of retrieving information connected to the CSF variation in the case of a generic dementia-related disorder by making use of fast optical non-invasive measurements. Specifically, we will show how time-resolved Near Infra-Red (tr-NIR) measurements have the potential to extract information about the presence of the disease, being effectively capable of monitoring its development. Carrying on such kind of studies requires careful design and important investments that must be evaluated in advance, in order to estimate the impact of the proposed research. To do to, we propose a computational approach to study the visibility of the dementia evolution via optical techniques. We will firstly approach the schematization of brain atrophy progression (BAP) in two geometries: a simple cylindrical model with parallel layers and a more complex human-head atlas. Then we make use of Monte Carlo (MC) photon propagation technique to simulate the results of a hypothetical longitudinal experiment on a human being. The goal is to exploit modern optical imaging techniques aiming at the definition of a novel way to monitor dementia diseases, with less expensive and non-invasive laser-based measurements. Explicitly, MC simulations are designed to replicate exactly a tr-NIR measurement, in which photons are probed into the human body and their reflected signal is detected from the same side. In such a way, it is possible to estimate the distribution of time of flight (DTOF), which carries important information about the photon diffusion within the underlying tissues. From the results obtained by running MC simulations while changing the brain volume, we expect to appreciate and characterize variations in the DTOF, thus linking these features to the disease progression.

Although many types of dementia imply brain atrophy, we will refer to the specific case of the AD in which the whole brain shrinks due to a surface erosion mechanism. The same results could be extended to study other neurodegenerative diseases by taking into account more specific brain modifications, such as for example FTD and NPH. Furthermore, these studies could be of interest also for the detection and monitoring of hydrocephalus, where recent reports have demonstrated the possibility to detect such alterations by frequency-domain NIR (fd-NIR) measurements. Potential for assessing these alterations and monitoring drug therapy in low income countries is particularly fascinating [18]. Our work takes inspiration on recent computational studies on the role of CSF in photon propagation [19] [20] [21] and novel understandings in AD progress prediction [22].

# STRATEGIES TO MODEL BRAIN ATROPHY

To mimic the disease evolution, we realized two different structural models making use of the mesh generator software iso2mesh [23]. As we already mentioned before, we focus our analysis mainly on dementia diseases that imply generalized neuronal atrophy affecting the external surface of the brain, leaving more specific structural effects as a further discussion. Without losing generality we will refer to this kind of degeneration as an ideal AD progression [24], which represents a broad 80% of the total cases of diagnosed dementia in patients.

## Brain atrophy progression assumption:

In the current literature, no structural meshes are available to model the BAP of any dementia disorder. To compensate for the lack of models, we decided to create a set of atlases mimicking the AD progression. The creation of the models is grounded to a simple assumption that is also one of the major candidate indicators of the diseases itself: as the disease progresses, the brain starts losing neuronal tissues and its volume shrinks [16] [17]. The effect associated with such tissue loss is that the grooves start to widen and the ridges get narrower [25], as if the tissue were eroded from its outer surface. Remarkably, the human body contrasts this cerebral-loss by filling up the empty volume left in the subarachnoid space with more CSF, increasing the thickness of the region where the fluid circulates. This is outright visible in MRI or XCT scans of the human head affected by the various kind of dementia [2] [25] and promotes the CSF as one of the possible biomarkers indicating the presence of the disease.

## BAP in a Cylinder model:

The use of layered samples that have optical properties similar to the tissues of interest is a simple approach to study the photon diffusion within a biological tissue. For this purpose, we created several 4-layered cylindrical mesh volumes, considering each layer as a different tissue: *Skin and Skull* (SS) as a unique entity, CSF, *Gray Matter* (GM) and *White Matter* (WM). In each mesh, the SS and GM tissues always have the same thickness, while the CSF increments in size to mimic the AD progression. The WM tissue was decreased to keep constant the total height of the cylinder, set to be 60 mm. We assume that no photons will trespass the cylinder, due to the very high scattering properties of the WM region. In total, we considered seven CSF thicknesses ranging from 0.0 mm to 15.0 mm in steps of 2.5 mm, which are shown at a glance in the infographic of Figure 1. On top of the SS surface we inserted 280 nodes, arranged in a concentric fashion. Each node defines the location of a time-resolved photon detector, forming 7 ring-arrays constituted by 40 photon counters. The distance between each ring is increased in step of 5 mm, having interfiber distances ranging from 10 mm to 45 mm. Each detector has a radius of photon acceptance of 0.75 mm and a total recording time window of 6 ns from the moment of the launch. A pencil beam source impinges in the center of the mesh and the resulting detection response functions are stored for each of the different meshes. The results obtained from all the detectors located at the same S-D distances can be integrated –due to the bi-dimensional isotropicity of the cylindrical model– increasing the signal to noise ratio by a factor of 40. This is a common averaging approach used in simple geometries such as in Monte Carlo Multi Layered (MCML) [26].

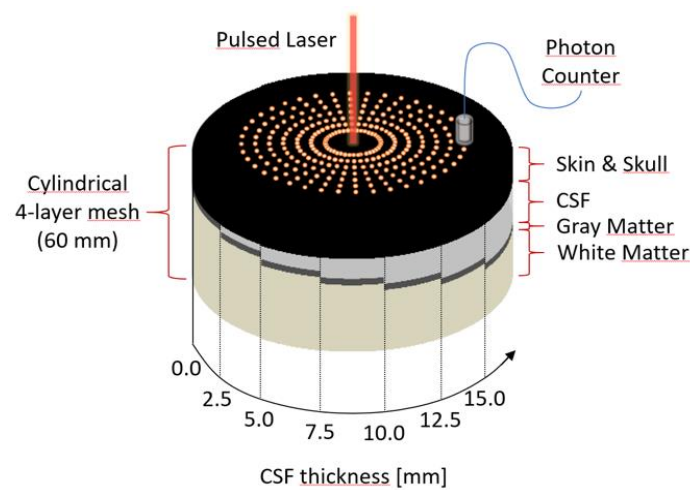

**Figure. 1 |** Infographic showing a combined view of various cylindrical 4-layer mesh volumes used in the simulations and the locations of source and detectors on its top layer. Preserving the Skin & Skull and the Gray Matter thickness, a linearly increasing CSF layer is inserted in between, representing a hypothetical Alzheimer's progression (or, in general, any brain atrophy dementia-related disorder).

**Table. 1 |** Tissues optical properties used in the simulations. The same tissue optical properties are used for both the cylindrical and the human head model. The last column shows the thickness of the layers used in the Cylindrical Model. For the CSF we examined two different scattering coefficients in this study.

| # ID | Tissue | $\mu_a$ $[mm^{-1}]$ | $\mu_s$ $[mm^{-1}]$ | $g$ | $n$ | Thickness (Cylindrical Model) $T_{ID}$ $[mm]$ |
|------|--------|---------------------|---------------------|-----|-----|-----------------------------------------------|
| **SS** | Skin and Skull | 0.019 | 7.800 | 0.89 | 1.40 | 12.5 |
| **CSF** | Cerebral Spinal Fluid | 0.004 | 0.009 [27] 0.1 [28] | 0.89 | 1.40 | Variable $\in [0.0, 15.0]$ step variation = 2.5 |
| **GM** | Gray Matter | 0.020 | 9.000 | 0.89 | 1.40 | 4.0 |
| **WM** | White Matter | 0.080 | 40.900 | 0.84 | 1.40 | $60.0 - (T_{SS} + T_{CSF} + T_{GM})$ |

## BAP in a realistic Human Head model:

Although planar layered phantoms are commonly used to compare and test experimental measurements, the availability of more accurate models can be useful to understand the role of geometrical structures in the photon propagation [29] [21]. Currently, one of the most accurate human-head mesh structure is the Colin27 model [30] [31], composed by a 4-tissues atlas. On the other hand, nothing similar exists for modelling a patient that exhibits BAP at different temporal stages. The lack of such interesting models is compensated in our work by the creation of several meshes progressively reducing the brain dimensions. Since dementia-related phenomena evolve by eroding the brain structure, we decided to perform this operation on both the WM and GM regions of the Colin27 mesh. Firstly, we separated the four tissues of the model, neglecting the inner closed surfaces in the brain that represent the ventricular cavities. For our specific purpose, there is no hope to detect photons reflected from such deep structures. The surface meshes extracted are enclosing each other with the same order as in the cylindrical model: starting from the outer layer toward the inner one we find the SS, the CSF, the GM and the WM. Each closed-surface was converted into a binary volume, with isotropic voxel resolution of 0.02 mm in the three spatial dimensions, then the enclosed regions were filled. The result of this operation returns four volumetric datasets containing the atlases of each structure considered. Two of them were left unaltered, SS and CSF regions, since the external shape of the scalp and the inner surface of the skull are not structurally affected by the AD disease. Every modification that follows was performed on both GM and WM binary datasets. Since the disease progression is associated with the loss of brain matter, we decided to operate exactly in the same way the AD acts aiming at miming the nature itself of the disease.

A binary erosion-operator [32] was applied to both GM and WM volumes, eroding the tissues with different structuring elements. Using first neighborhood erosion (cubic and diamond structuring elements in Figure 2) resulted a bad choice: single point cavities in the brain, such as small foldings and grooves, started to develop into artifacts similar to the structuring element used. The top panel of Figure 2 shows what happens by eroding with such structures, where squared or triangular cavities start to be clearly visible at later erosion stages. Instead, using a spherical-like structuring element gave smoother results, preserving the original curvature of the foldings at the exterior brain surface (Fig. 2, bottom part). In this preliminary study, we consider the latter to be a good approximation to model the brain erosion mechanism behind dementia diseases.

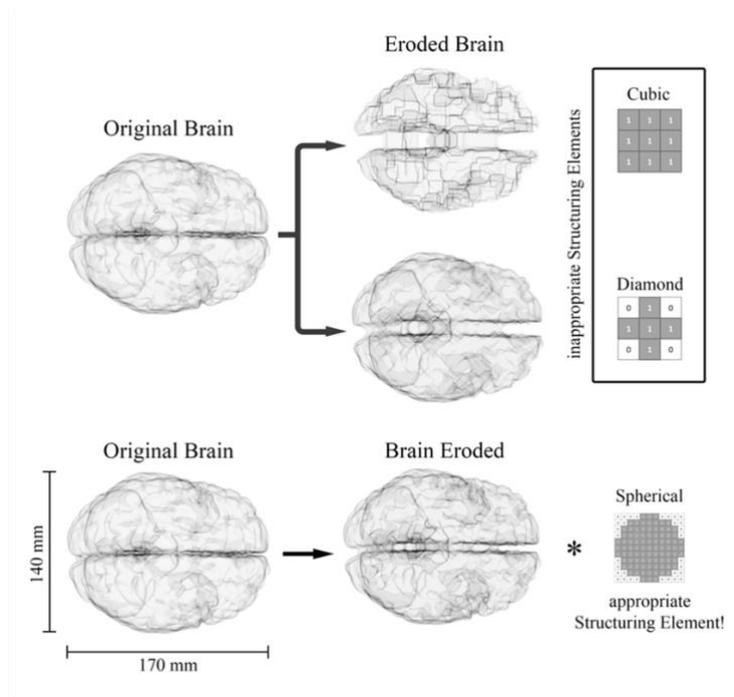

**Figure. 2 |** The brain erosion process in voxel coordinates. On the top, it is possible to notice how, using cubic or diamond structuring elements results in a developing of artifact as soon as the brain shrinks. More realistic instead, an erosion modeled with a nearly-spherical element (bottom), which leads to

rounded erosion and does not turn into shape artifacts. With the ∗ operator we define the dilation, inverse of the erosion operation, which once applied to the eroded brain would return the original version. For graphical purposes we describe a 2D structuring element (right column), while in our work we have effectively used its 3D counterpart.

We point out that it is not possible to proceed stepwise (1 voxel erosion at each step) and the diameter in the spherical erosion core has to be greater than D = 4 vx. In fact, choosing D = 1 vx or 2 vx gives rise to cubic artifacts and D = 3 vx has diamond evolution. In general, the larger the D of the spherical structuring element, the smoother will be the erosion. In our work, we used erosion with spherical structuring element of diameter D = 6 vx = 1.2 mm generating 11 meshes that mimic hypothetical stages of the disease. The whole process took around 2-5 minutes per mesh with a normal PC equipped with a CPU intel i7-4930K and 32 Gb of RAM depending on the BAP stage. More advanced stages have less nodes on the GM and WM surfaces and the meshing time decreases.

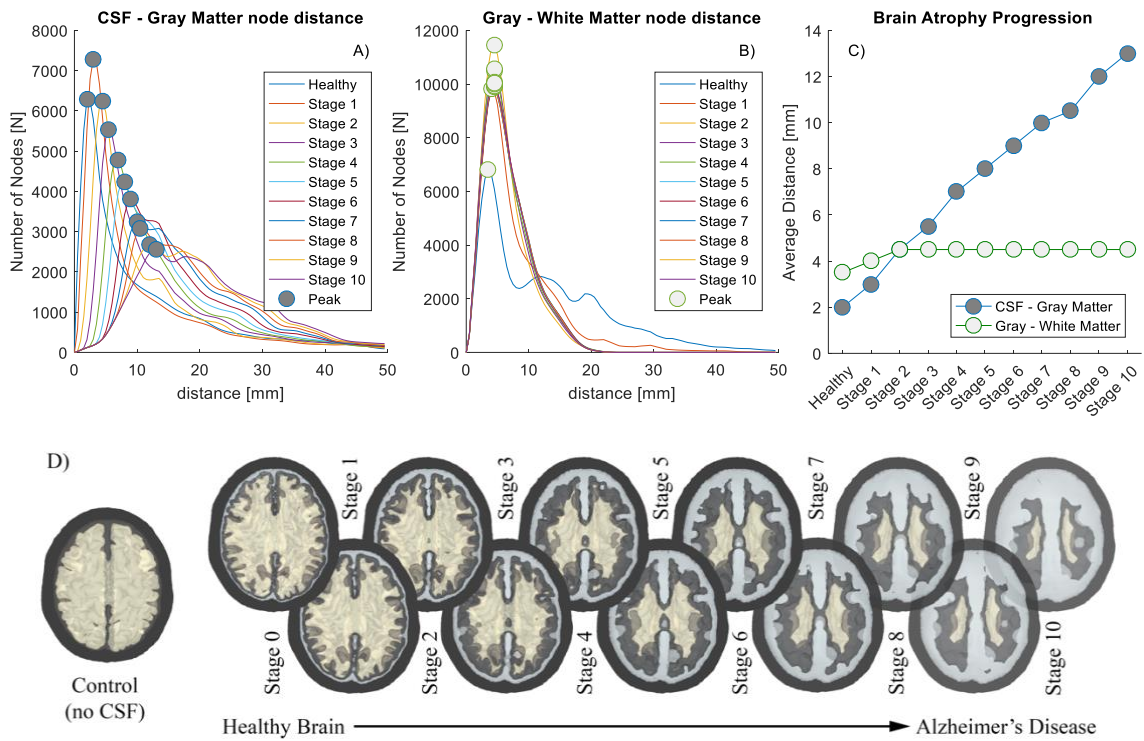

**Figure. 3 |** Informative about the Alzheimer's model creation. From the top graphs in panels A-C, it is possible to notice how the shrinkage of the brain is linear with respect the increasing thickness of the resulting CSF layer. Both the GM and the WM were shrunk independently with the same criteria, which results to preserve their average distance (and so the average thickness of the GM). D) The bottom infographics shows a tomographic cut of the models and their respective name used throughout the text.

Once the AD models are created we compare their average geometrical features with those of the cylindrical phantoms, trying to be consistent with a real brain shrinking in human. Unfortunately, at the best of our knowledge, the literature is limited and there are no systematic studies on dementia progression that allow accurate quantitative comparison. On the other hand, a pretty variegated literature on structural brain imaging offers a rough estimation of the extents of the BAP in clinical cases. Here we report some analysis, assuming all the patients having the same head size of Colin27 (reported in Figure 4). In this way we are able to calculate the distance of the brain surface with respect to the inner part of the skull, obtaining an estimation of CSF thickness in different zones affected by atrophy. As an interesting example, we report the work of Yi et al. [25] that analyses several different patients affected by AD or FTD (Figure 1), ordering them in function of their BAP stage. Units are not displayed, but we can estimate the CSF thickening up to ~10 mm at stage 4. Another interesting example is reported in the review from Harper et al. [2], where they show in Figure 4 various forms of brain atrophy (although they do not insert any healthy control as a reference). In this case we can take advantage of a patient showing asymmetric atrophy (Figure 4, 3rd image from the left, top row) clearly showing a right hemisphere strongly eroded if compared to the left one. In this case, the severe right hemisphere shrinkage was in the order of ~12 mm. In the next patient (Fig. 4, 4th image from the left, top row) showing parietal/occipital atrophy, the parietal lobe is ~9 mm apart with respect the skull. Unaffected by atrophy, the temporal lobe is distant a few mm in agreement with what we measured in *Stage 0* of our models. An extreme example, such as the case reported by DeBrito-Marques [33], shows a patient having amyotrophic lateral sclerosis with dementia (Fig. 2), exhibiting an impressive temporal lobe atrophy. Measurements are not displayed, but the normalization of the head dimensions let us estimate that the brain shrinkage was 30-40 mm. Such impressive atrophy was localized in the temporal lobe, the region of the brain that we are going to study in the present work. This let us believe that erosions in the range of ~10 mm still represents a clinical case of interest.

In total, we have built 11 meshes: *Stage 1* to *Stage 10* for the AD and *Stage 0* for the original Colin27 model. As a plus, we consider another mesh that we call *Control,* obtained by removing the CSF layer (substituted with GM) and dilating the WM one step (D = 6 vx). The results of the erosion process and successive re-meshing are presented in Figure 3 panel D together with their geometric characteristics. Per each surface, we calculated the nodal distance probability distribution (NDPD) of each couple of enclosing layers (SS surface encloses CSF, CSF encloses GM and GM encloses WM). Per each node of the inner surface we find the minimum distance from the outer surface, allowing us to build the NDPD histogram (Figure 3 panels A-B). Considering its maximum as the average distance between two surfaces, it is possible to estimate the mean thickness of every tissue considered. In agreement with the *Cylindrical Model,* the nodal distance between SS and CSF surfaces is $T_{SS} = 14.5 \pm 1.5\ mm$ and it represents the average thickness of the skin and skull tissue that we assume unaltered in every AD Stages. From the plots of Figure 3 results clear that the average thickness of the GM is preserved during the erosion, while the CSF's increases approximately in a linear fashion. Once the CSF thickness is characterized at every *Stage*, it is possible to simulate the time-resolved photon propagation using the tissue properties reported in Table 1. We define a pulsed beam source positioned on the right hemisphere, as show in Figure 4 panel A), pointing perpendicularly to the SS surface. The underlying cerebral region corresponds to the temporal lobe of the right hemisphere, like in the clinical case of [33]. Four time-resolved photon detectors were located at 10 mm distance one respect to the other. Figure 4 clarifies the position of the source and detectors on the human head and helps visualizing the reduction of the brain volume of the AD models. In such figure, the left GM brain hemisphere at Stage 0 (Colin27) and the right one at Stage 6 are rendered together, visually resembling the clinical case of asymmetric atrophy [2].

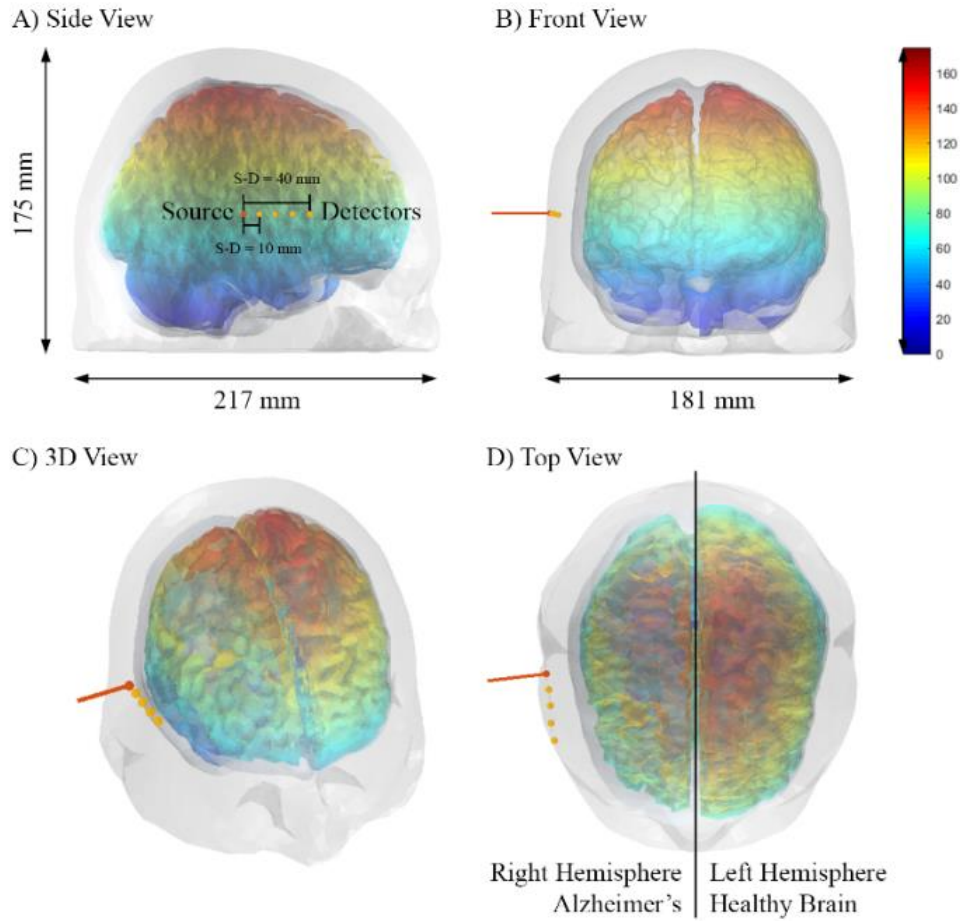

**Figure. 4 |** The pulsed laser source impinging the human head from its right hemisphere. The detector positions are shown in orange and their distance is 10 mm with respect each other. In this figure, the right lobe of the brain is from the AD model *Stage 6* while the left is the original Colin27 model [31]. To help the visualization of the modeled disease progression, we label the $z$-coordinate with jet color bar.

## Measurement scheme and Instrument Response Function

To take into account the perturbation introduced by a realistic measuring system, we acquired experimentally the Instrument Response Function (IRF) of a typical detector used in fNIRS laboratories [34]. We used a picosecond pulsed laser operating at 830 nm (LDH-P, SEPIA II, PicoQuant GmbH, Germany) and a hybrid detector (HPM-100-50, Becker & Hickl GmbH, Germany). The detector was coupled to a Time-Correlated Single-Photon Counting board (TCSPC) (SPC131, Becker & Hickl, Germany), using a full scale of 4096 channels and a resolution factor of 3.05 ps/channel. The measured IRF is reported in Figure 5 and it is convoluted with the simulated MC photon distributions to analyze the measurement effect on a real system. The IRF used for the convolution was cut, normalized and rescaled to a total of 40 channels with time step of 40 ps/channel, leading to the final IRF shown in the nested plot of Figure 5.

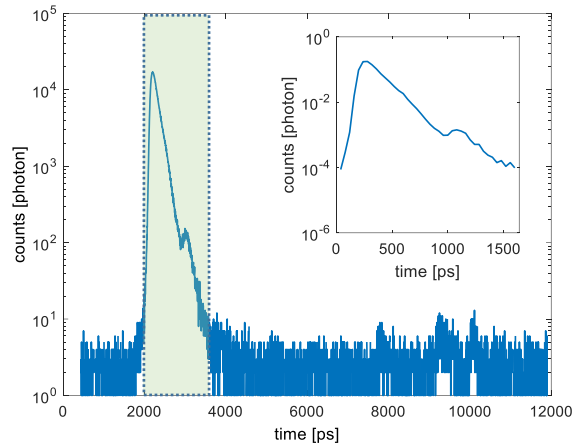

**Figure. 5 |** IRF used to reproduce realistic laboratory measurements. The main plot shows the measured IRF of a common detector system and on the nested plot its extracted and normalized version (that corresponds to the green region in the main graph) that we used in our study to reproduce the effect of a realistic measurement.

## RESULTS

In general, light transport within any biological tissue is mathematically described by the Radiative Transport Equation (RTE) [35]. Exact analytical solutions for the RTE are accessible only with simple geometries or with Diffusion Equation (DE) approximation [36], whereas for complex shapes other approaches should be taken into account to correctly infer the photon propagation. Among the others, the Monte Carlo (MC) photon propagation method statistically provides an accurate solution for the RTE when high number of photons are propagates [37]. Because of this, it is considered the gold-standard method for modelling the photon diffusion within biological tissues and it is currently widely used as a predictive tool in the optical imaging community [28] [38] [39].

Thanks to modern paradigms in CPU and GPU computing, Monte Carlo (MC) simulations [40] [27] are becoming a feasible choice for modelling the photon diffusion within biological tissues. In this study, we took advantage of one of the most recent and accurate Monte Carlo simulation tool for the photon propagation, the Mesh-based Monte Carlo (MMC) [27], to resolve the photon diffusion within the human-head model. A total of $1.1 \cdot 10^{10}$ (eleven billion) photons were launched, with an average running time of 10 hours to propagate $10^9$ (one billion) photons per each AD models. For the scattering coefficient of the CSF we studied the effect of two different values as reported in Table 1, differing by one order of magnitude. The values were taken from relevant bibliography [27] [28] and represent two extreme cases. In the following, we analyze the results obtained for the *Cylindrical* and *Human Head* meshes, focusing in particular on the possibility of following the disease evolution in time in a realistic laboratory measurement scenario.

## BAP in the Cylindrical Model

Already with the cylindrical model for the AD we obtained interesting results from the photon diffusion within the phantom. In Figure 6, we plot the distribution of time of flight (DTOF) –also called time-resolved diffuse reflectance– measured at various interfiber separations. Although we sampled also at other distances, we do not plot the results at intermediate values (15 mm, 25 mm, 35 mm) since they would be visually redundant and do not add further information. It is visible from the graphs that increasing the CSF thickness does not turn into any interesting variation at early detection times. This is not surprising since the early photons have been spending most of their time in the superficial SS layer. This happens at short S-D separations (10 mm to 20 mm) due to the fact that the reflectance curves of the AD models share the same trend with the *Control* model without CSF. At longer distances (25 mm to 40 mm) instead, the AD and Control models start to diverge, suggesting contributions from deeper tissues. It is remarkable the fact that, at early times and at long S-D separations, all the AD models are not distinguishable between each other. This implies that the photons reaching the CSF levels delay their way back to the detection side, pushing the variations of the DTOF curve at middle-late timing.

Due to the above reasons, we do not appreciate any visible shift of the response peak (Figure 7, panel A), which variates at different S-D distances but does not between the ADs and Control models at given interfiber distance. All the DTOF at different CSF thicknesses are practically overlapping: a weak effect starts to appear at longer S-D distances that we do not consider relevant in this study. On the other hand, the effect of the convolution with the IRF is shown in panel B of Figure 7. In this case, the temporal shift is expressed as the delay in time of the convolved DTOF curve with respect to the IRF peak, located at 240 ps in the curve or Figure 6. Like in the ideal case, no perturbation is observed after the convolution while increasing the CSF thickness. The light-red regions in Figure 7 A-B represent the confidence band of the results obtained by considering a less scattering CSF layer.

At middle timing (1000 ps to 2500 ps), the DTOF response is found to be in an intermediate regime, in which photons reflected at a deeper level start to be seen from the detectors. More interesting, instead, is the situation arising from the analysis of late-timing response curve (>2500 ps). At such long times the photons have the ability to trespass the SS layer and reach the CSF. In such an optically clear environment, light propagates in a nearly-straight line and the time of flight increases proportionally to the size of the CSF. At late times, in fact, the variation of the size of the CSF turns into a change in the slope of the DTOF that loses its steepness while thickening the clear layer. This fact is evident in every plot of Figure 6. Panel C of Figure 7 reports the slope variation at late-timing due to the increased thickness of the CSF layer (solid line for strong scattering, dashed lines for weaker). We found a steep slope-change at short CSF variation for both the values of the scattering coefficient examined, which suggests higher sensitivity at early BAP. Remarkably, the detection at different S-D separations returns approximately a constant function slope, which we plot as a confidence gray band (and light-green for lower scattering). We conclude that this is a direct feature introduced by the CSF thickness and it can give hints about its inner structure.

The considerations about the slope measurement arising with a real IRF are even more interesting (Fig. 7 panel C, data presented with filled or empty circles). Even after the convolution e can appreciate that the slope of the response function still falls into the confidence band. The response of the measurement system, then, is not compromising the possibility to estimate the thickness of the CSF layer, since simulations and realistic measurements (affected by IRF) displayed the same trend.

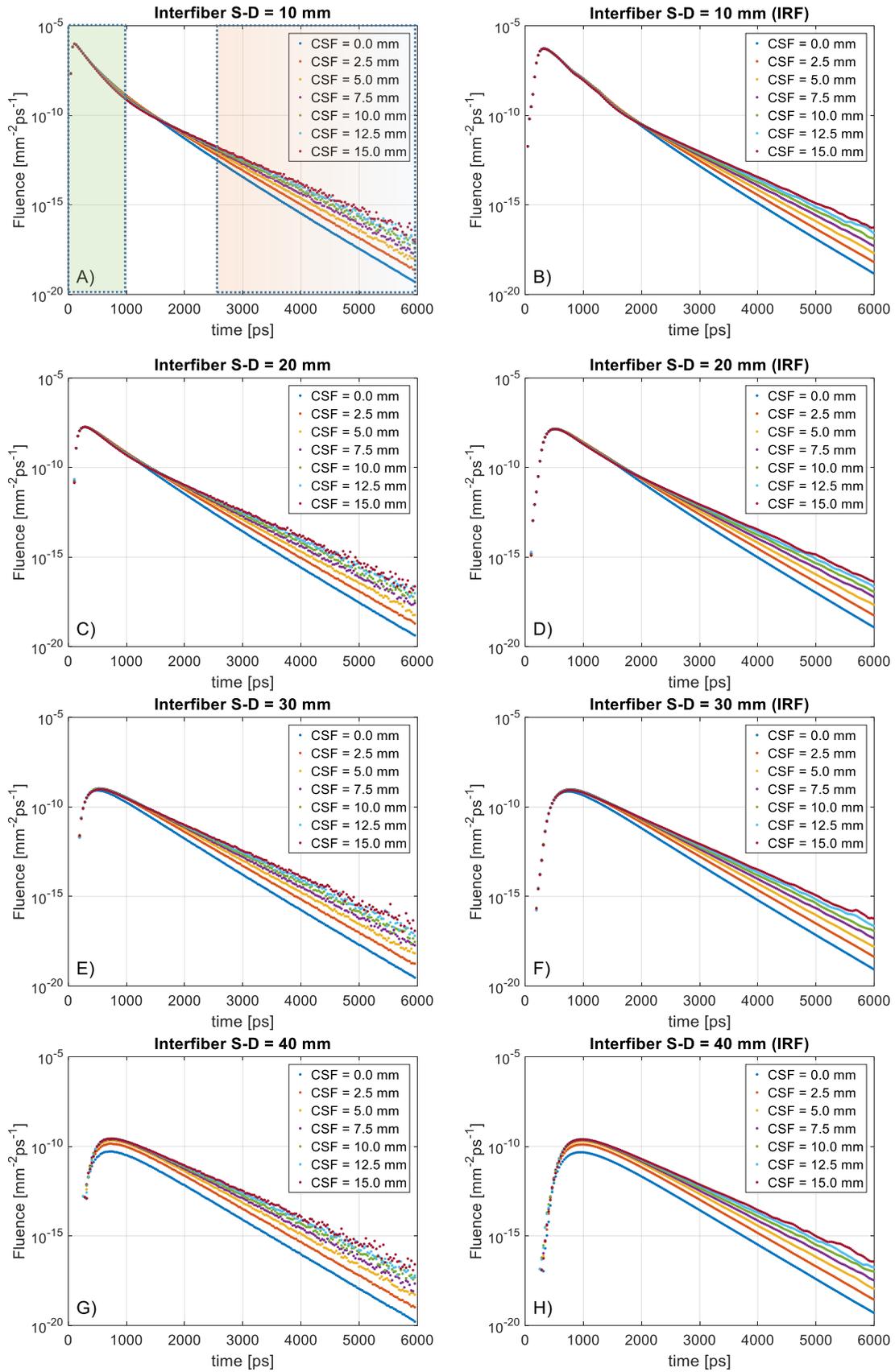

**Figure. 6 |** Distribution of time of flight (DTOF) at four interfiber source-detector (S-D) distances for the cylindrical layered phantoms with variable CSF thickness. On the left column, the raw data considering an ideal response of the detector, on the right the corresponding curves convolved with a typical instrument response function (IRF). In the first plot are marked the temporal windows considered as early-photons (green) and late-photons (red) gating.

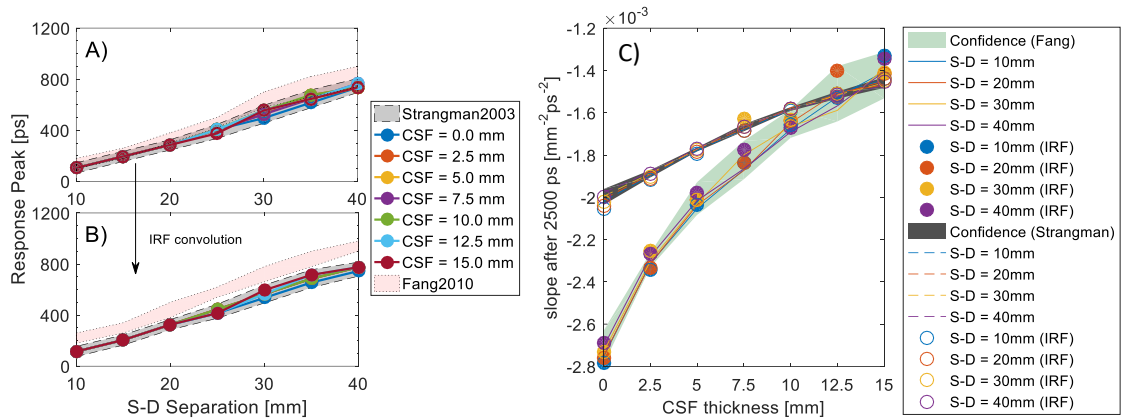

**Figure. 7 |** CSF fingerprints in the Cylindrical Model. On the left panel A), the peak response delays in function of the S-D separation and it resulted to be equal for all the CSF thicknesses investigated. On panel B, the corresponding response delay after the convolution with the IRF that introduces. On the right panel C), the changing in the slope of the late photons due to increased CSF thickness.

## BAP in the Human Head Model

So far, we examined the behavior of an ideal cylindrical layered volume under virtual tr-NIR experiment; but recent works in continuous wave regime [21], pointed out differences at detection level deriving from structural features of the clear layer. It is worth to examine any possible structural-induced effect also with time-domain measurements, comparing the results from the cylindrical model with those obtained with more detailed atlas.

We impinge the same laser source on the right head hemisphere as described in Figure 4, and we collect the detector responses in the same way we did for the cylindrical phantom. In this case, we cannot sum over detectors located at the same S-D separation due to the not symmetric shape of the human head. We performed the same simulation for every *AD model* created, but in Figure 8 we plot only the DTOF obtained for *Stage 0, 2, 4, 6, 8* and *10* to do not compromise the readability of the graphs. For the sake of completeness, we report that the intermediate results followed a continuous trend.

Surprisingly, the situation is pretty similar to what we obtained in the cylindrical models. Also in this case, there is no appreciable shift of the peak response at different CSF thickness, due to the fact that the photons contributing to early-time detections are travelling mostly in the superficial region of the SS tissue. As expected (Figure 9, panel A), increased S-D separation progressively delays the signal-peak while IRF-convolution does not introduce any effect on such delay (panel B). Similarly to what seen in the cylindrical model, the slopes of the DTOF are directly affected by the CSF thickness with approximately uniform sensitivity at every stage of the atrophy progression (Figure 8, panel C). Interestingly, one order of magnitude change in the optical properties of the clear layer does not variate the sensitivity of the slope-curve, which follows the same trend for both the $\mu_s$ analyzed (gray and light-green confidence band in panel C). Even if the general results are congruent with those obtained using cylindrical phantoms, the absolute values are different. Most likely this is due to the fact that the CSF region is now a complex surface with grooves and foldings, which definitely complicates the photon-path if compared to that of normal planar surfaces. Moreover, the relative size of the layers is approximately similar but not exactly matching, leading to different delays in the reflected signal. The convolution with the IRF turned into the same behavior described in the previous paragraph and does not influence the overall slope of the late-timing DTOF curve. This suggests that even in the more realist case of a human head, the chances to appreciate internal CSF variations are not compromised by using a real instrument.

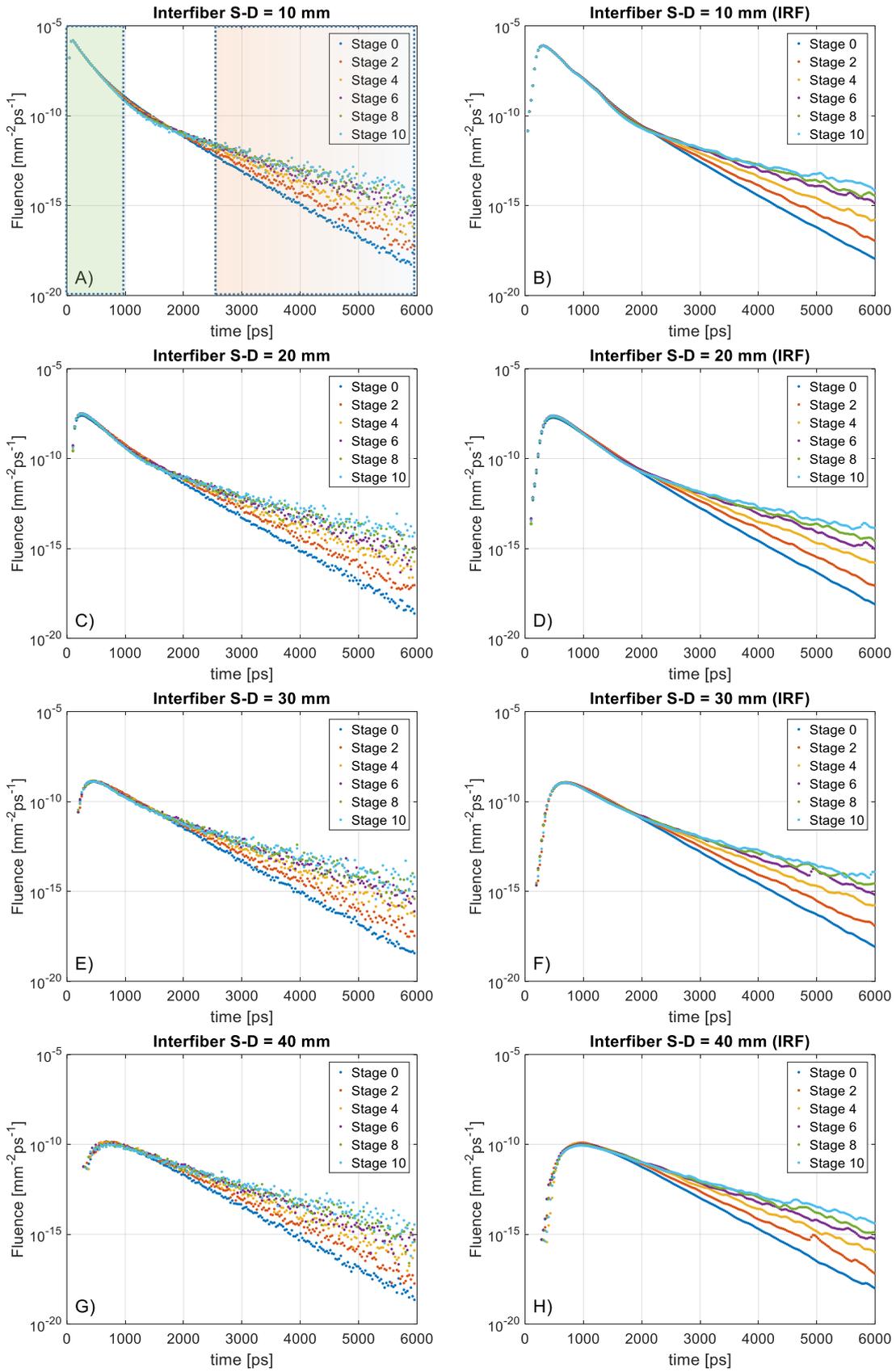

**Figure. 8 |** DTOF curves at various Source-Detector separations. Compared to that of *Cylindrical models*, the measurements are noisier due to the impossibility of averaging many detector responses because of the lack of symmetry of the human head model. The left column shows the raw datasets, on the right the corresponding convolution with a realistic IRF.

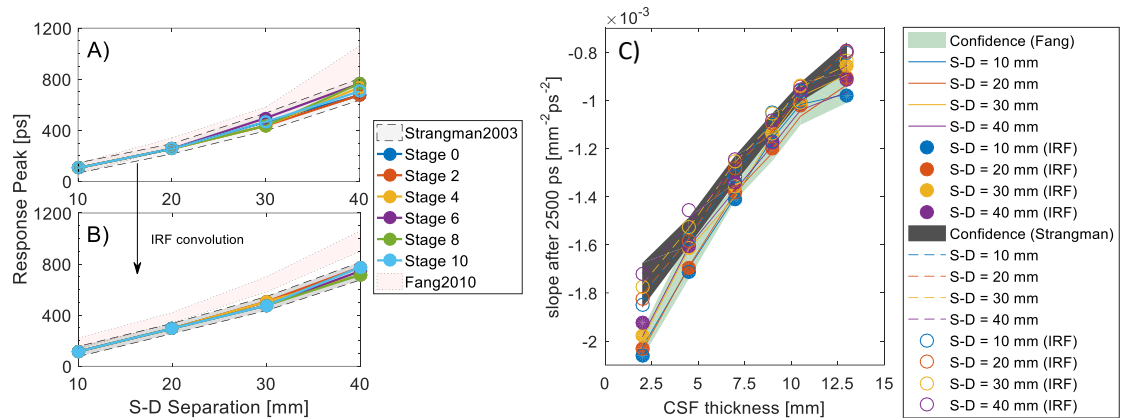

**Figure. 9 |** Features of the response curves for the Human Head models. A) There is no evident peak shift in the response curves at early detection times in function of the CSF. The peaks are so close to each other that in the plot all the curves are overlaying each other. B) Even the convolution with the IRF, which shifts up the peak response in time, does not introduce any further visible effect in the measurements. C) Slope variation for late-photons detection in function of the CSF thickness. In the plot is possible to notice how increased thicknesses of the transparent layer surrounding the brain will affect the reflectance curve, giving useful hints about the possibility to estimate its average thickness.

# DISCUSSION

As we already discussed throughout the text, AD and other dementia problematics generically involve variations of the brain-volume due to a neurodegenerative process that lead neurons to commit apoptosis. Although the biological mechanism is not yet clarified, such variations are always connected with a shrinkage of the GM and WM constituting the brain and, in particular, express themselves as a complex change on the brain cerebral cortex. This surface erosion implies specific modifications, which turn into a widening of the grooves and narrowing or the ridges typical of the neuroanatomical structure of the brain. Ventricular cavities are also affected by the neuron-loss connected with the disease, leading them to an expansion of their chamber-volume. During the disease progression, the brain is effectively reducing its size inside a fixed braincase and the space left-over is filled with an increased amount of CSF. This is a physiological reaction that tries to preserve the mechanical and immunological functionalities of the CSF that surrounds the diseased brain. From a structural point of view, the brain changes its relative mean surface-distance with respect the internal skull-walls, since the neurocranium does not variate appreciably in shape (especially in normal elderly individual). This is functional to the progression of the neurodegenerative process and it is an important parameter for assessing the health condition of a patient. At the moment, the techniques that could give structural information about the dementia progression are limited to the costly and non-portable MRI or to the radiation-exposing XCT. Despite their adequate imaging resolution for the detection of structural changes, time-resolved monitoring of the disease is not yet affordable due to high risks connected with repetitive ionizing radiation or –more in general– with the high cost that such study would imply.

In our work, we tried to tackle the current blindness in longitudinal studies of structural disease evolution by exploring the possibility offered by harmless tr-NIR measurements. To understand the role played by the CSF in the DTOF response curve, we created ad-hoc detailed mesh-models that replicate the structural disease evolution in time. The absolute lack of systematic studies for such evolution was compensated by the creations of different cylindrical and human-head meshes that model the disease progression in time. A simplistic study in cylindrical geometries with varying CSF thickness was complemented with the reproduction of an AD model that realistically mimics the brain atrophy, in order to analyze the role of complex structural conformations. To create these detailed *Human-Head models* we took advantage of modern meshing tools [31], which we exploited to implement an erosion algorithm reproducing the brain-shrinkage. We decided to use the Colin27 model [30] and then erode the gray and white matter via an iterative process, accomplished converting the mesh into voxelized atlases and re-meshing them to obtain the final models. With these new *Human-Head* structures we do not have the ambition to represent an exact model of the AD development: they offer the opportunity to predict a general trend of realistic tr-NIR measurements, keeping an eye at the nature of the BAP. In fact, our erosion method preserves the complex structure of the brain surface, with grooves and ridges evolving similarly to the neurodegeneration in a real individual affected by dementia [2] [25] [33]. Of course, further details could have been taken into account, such as the modelling of the ventricle size variation, but at the first step of our work there is already a broad spectrum of useful information arising from this study.

Here we tried to obtain general results, parametrizing the CSF with two different scattering coefficients: firstly considering it as a nearly transparent and non-scattering layer [27] then having more turbid optical properties [28].

A great number of MC simulations of ideal tr-NIR experiments were run during this study. Each virtual experiment was composed of 10 partial simulations obtained propagating $10^8$ photons, in order to have a sufficient number of DTOFs for statistical analysis. The standard deviation of each curve in Figure 6 and 8 was divided by their average, obtaining the coefficient of noise-variation of the MC simulation. In Figure 10 we report the analysis of the curves measured at 30 mm interfiber distance with Strangman's parameters, but similar results were obtained at all the S-D separations. In the cylindrical model (panel A) all the simulations are affected by an error that is always smaller than 5%, while in the human head model (panel B) the error gets bigger than 5% after 4000 ps. Such a small statistical noise gives enough accuracy for calculating the DTOF slope between 2500 ps and 4000 ps. Interestingly, changing the CSF does not add any visible effect in the noise-trend that seems to be related only to pure measurement conditions rather than geometrical variation. This drives us to the assumption that the MC simulations have the same sensitivity at each BAP stage and all the results that we will discuss in the following are comparable with each other.

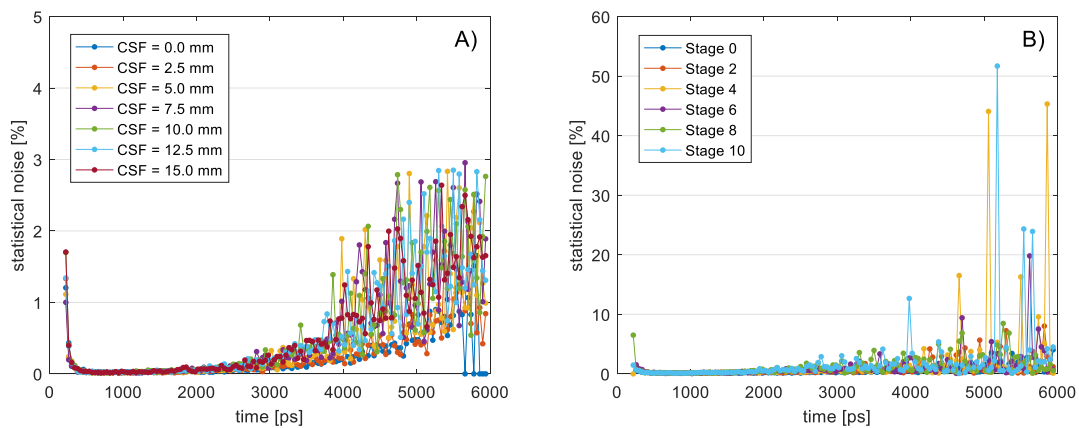

On the other hand, the DTOF curves calculated with the detectors resulted to be sensitive to CSF thickening, but interestingly only few features were influenced by the change of the clear layer. Firstly, we found that increased level of CSF does not introduce any significant peak-shift of the DTOF (Figures 6-8). Moreover, at early times, all the curves with different CSF thicknesses are very similar one to the other making impossible to distinguish between them (Figures 7-9 panels A). Not even the convolution with an experimental IRF modified this trend, thus suggesting that a real measurement would not introduce any unwanted artifact (Figures 7-9 panels B). These facts indicate that the disease evolution cannot be seen at early photon gating, because the light did not sufficiently explore deeper tissue regions. Of more interest is the analysis of late-photons (starting at around 2500 ps) where the DTOF starts to exhibit a linear decay (in logarithmic scale, Figures 6-8). In this case, the shrinking of the brain region visibly changes the slope of the detector response (Figures 7-9 panels C). Higher thicknesses of the clear layer correspond to less steep slopes, suggesting that following its variation could help monitoring the disease evolution or the assessment of its stage. More interestingly, this effect seems not to be affected by the IRF convolution that exactly preserved the slopes (Figures 6-7 panels A-B), further increasing the chances to use it as biomarker in realistic laboratory measurements. It is worth to notice that the slope –at late gating– does not variate at different interfiber S-D separations, which gives the opportunity to consider averaging different response curves in case of noisy measurements.

Remarkably both cylindrical and human models behaved in a similar fashion with respect to the slope variation: increasing the thickness of the CSF flattens the slope of the DTOF at late time gating. On the other hand, our previous work on Fluorescence Molecular Tomography [41] shown that the size of the CSF had no role in the photon diffusion process in a cylindrical model. In that case, the models were insensitive to thickness variation, up to the point that the transparent layer could have been neglected without losing accuracy. On the other hand, in a more realistic model [42], the inclusion of the CSF affects the simulation outputs, turning into output artifacts that are not possible to normalize [21]. Depending on the geometry of the model and the measurement scheme employed, clear tissues exhibit different behaviors that have to be singularly evaluated.

In this study, the trend followed was the same for both the models and the optical parameters: the DTOF after 2500 ps followed a linear decay, decreasing its slope steepness while increasing the CSF volume. On average, the absolute difference of the slope variations between the models was around 35%, suggesting that the models were sensitive to the CSF's structure and to its optical parameters. To analyze the statistical significance of this difference in terms of curve's trend, we performed T-tests on the first derivatives of the plots in Fig.7-9. Setting a significance threshold of p=0.05, the test was satisfied for every curve comparison (dashed lines in Fig. 11) except for the cylindrical model with Strangman's parameters (red, green and blue solid lines in Fig. 11). In this case, the slope change exhibited a weaker growing with respect the other situations and, as a general trend, we found that the T-test was better satisfied while increasing the CSF thickness. This result suggests that at higher depth the structural details progressively lose importance in the photon propagation. For the sake of completeness, we report that the T-tests were never satisfied while comparing directly the slope-change curves, due to the different absolute values. The fact that the trends are the similar under different conditions is important for the robustness of the technique. In fact, structure and tissue optical properties could sensitively change during the lifetime of an individual.

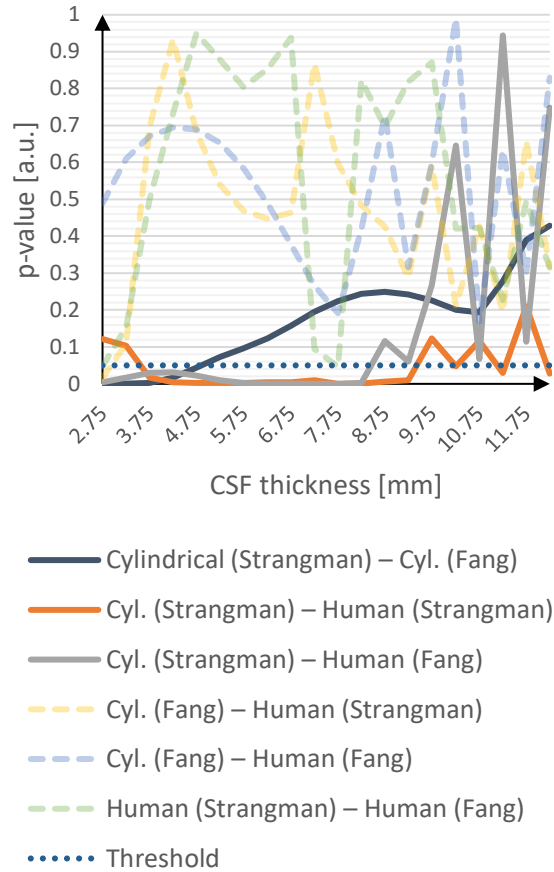

**Figure. 11 |** Comparison T-tests on first derivatives of the slope variation curves. The plot reports the p-values of the null hypothesis of the curves having the same trend. When the values are greater than the threshold 0.05 the models compared have approximately the same slope variation.

At this stage, the results clearly demonstrate a significant effect caused by the progressive brain atrophy on the shape of the DTOF. Whether these alterations are detectable or distinguishable from physiological variations in healthy subjects, is yet to be addressed. It is out of the scope of the present study to investigate effective strategies to disentangle healthy from diseased subjects, or at least to longitudinally monitor the very same subject over time. In principle, as compared to other locations as the forearm or the abdomen, the head geometry is rather fixed over time and chances to track internal changes could be viable following the approach we proposed.

# CONCLUSIONS

Some important achievements emerged from this study. For the first time we approached the problem of the creation of complex human head atlases with variable CSF thickness, mimicking biological tissue variations comparable with that of dementia related diseases. Although complex, the protocol that we followed is quite linear and could be also tuned to describe some more specific brain variations, such as temporal-lobe atrophy (in the FTD) or enlarging ventricular cavities (in the NPH). Moreover, MC photon propagation seems to be the natural choice to perform such study: other approaches would have neglected the presence of the CSF layer [43] or approximated it with radiosity theory [44], inserting unpredictable perturbations to the model. Instead, the MC is often considered as a gold standard for the simulation of the photon flux in biological tissues and let us retrieve important information about the effect of CSF in a realistic tr-NIR measurement. Late photons, shown an interesting slope-change in the DTOF that does not depend on the interfiber distance but rather on the CSF thickness itself. If accurately measured with longitudinal studies, such slope change could give precious information about the disease stage and the velocity of its evolution, thus opening new paths towards the understanding of the disease mechanisms. On the other hand, late photon gating (>2500 ps) might be difficult to achieve in real systems with acceptable SNR, but progresses in the creation of more efficient detectors are pushing this limit further towards late photon analysis [38]. At the moment we notice that the convolution with an experimental IRF did not influence the results, thus promising a certain degree of accuracy retained also at experimental level. Of course, further investigations could be done by taking into account other IRFs of various detectors, but this would have exceeded the purpose of this study. Lastly, results obtained with a realistic head model and with a simpler layered structure were comparable, showing the same trend for the slope-change effect at late gating. Even if quantitatively differing from each other, this reinforces our belief that slope variation is tightly connected with the average CSF thickness rather than local structural conformations.

All these considerations leave us with a promising perspective for the real implementation of this procedure, making our results more general and weakly depending upon exact anatomical surface features. To extract such details in fact, MRI and XCT measurement are still the gold-standard, but hints regarding the CSF thickness-change are likely to be captured also by faster and non-invasive tr-NIR measurements. Although our study is broad, a lot more could be done in the near future, to further investigate the behavior found in our simulations. Not only real experimental tests are on our future plan, but also new simulations are on the way to test different and possibly more accurate methodologies. Among others, tr-NIR Spectroscopy (tr-NIRS) is a major candidate from which we could extract further information in function of the wavelength. In particular, seems interesting the exploitation of the complex absorption landscape of the water to obtain a reference curve independent on the CSF thickness. Our goal, in fact, is to open up the possibility of tr-NIRS imaging for neurodegenerative monitoring. For these reasons, we believe our contribution to the biomedical imaging field to be of renewed interest for the dementia-research community. With this work we hope to foster novel strategies for the definition of non-invasive biomarkers, aiming at a better understanding of the dementia diseases and -possibly- towards a better care of the patient.

**AUTHOR CONTRIBUTIONS**

D.A. conceived the idea and defined the research path with A.P. D.A. designed the models and run test simulations. D.A., G.Z., L.S., A.T., A.P. conceptually designed the computational experiment and discussed the results. L.Q. run the simulations, participated in all the discussions and extracted the numerical results. D.A. and L.Q. analyzed the results and discussed possible further investigations. D.A. wrote the manuscript and made the figures, all the authors contributed to correct the manuscript. D.A. friendly thanks Dr. Vincenzo Lagani for helpful suggestions concerning statistical analysis of the datasets.

**DISCLOSURES**

All of the authors declare to have no relevant conflicts of interest.

**ACKNOWLEDGMENTS**

This work was supported by the grants "Skin-DOCTor" (1778) implemented under the "ARISTEIA" Action of the "Operational Programme Education and Lifelong Learning" co-funded by the European Social Fund (ESF) and National Resources, the EU Marie Curie ITN "OILTEBIA" PITN-GA-2012-317526 and the H2020 Laserlab Europe (EC-GA 654148).